\documentclass[english,aps,prl,reprint]{revtex4-1}

\usepackage[english]{babel}
\usepackage{amsmath}
\usepackage{amsthm}
\usepackage{amsfonts}
\usepackage{amssymb}
\usepackage{amscd}
\usepackage{amsbsy}
\usepackage{graphicx}
\usepackage{booktabs}
\usepackage{array}
\usepackage{subfigure}
\usepackage{enumerate}
\usepackage{url}
\usepackage[nodayofweek]{datetime}
\usepackage[english]{babel}
\usepackage{mathtools}
\usepackage[toc,page]{appendix}
\usepackage{verbatim} 
\usepackage{amsthm} 
\usepackage{enumitem}
\usepackage{enumerate}
\usepackage{bbm} 
\usepackage[normalem]{ulem} 
\usepackage{bm}
\usepackage{mathtools}

\newcommand{\be}{\begin}
\newcommand{\e}{\end}
\newcommand{\beq}{\begin{equation}}
\newcommand{\eeq}{\end{equation}}
\newcommand{\beqs}{\begin{equation*}}
\newcommand{\eeqs}{\end{equation*}}
\newcommand{\bal}{\begin{align}}
\newcommand{\eal}{\end{align}}
\newcommand{\bals}{\begin{align*}}
\newcommand{\eals}{\end{align*}}

\renewcommand{\l}{\left}
\renewcommand{\r}{\right}
\renewcommand{\d}{\mathrm{d}} 

\newcommand{\curly}[1]{\mathcal{#1}}
\newcommand{\goth}[1]{\mathfrak{#1}}

\newcommand{\om}{\omega}

\newcommand{\lam}{\lambda}

\newcommand{\gam}{\gamma}
\newcommand{\Gam}{\Gamma}
\newcommand{\al}{\alpha}
\newcommand{\de}{\delta}

\newtheorem{thm}{Theorem}

\newtheorem{defn}{Definition}

\begin{document}

\title[Anomalous Lieb-Robinson Bounds]{New Anomalous Lieb-Robinson Bounds in Quasi-Periodic XY Chains}

\author{David Damanik}
\affiliation{Department of Mathematics, Rice University, Houston, TX 77005}
\author{Marius Lemm}
\affiliation{Department of Mathematics, California Institute of Technology, Pasadena, CA 91125}
\author{Milivoje Lukic}
\author{William Yessen}
\affiliation{Department of Mathematics, Rice University, Houston, TX 77005}



\begin{abstract}
We announce and sketch the rigorous proof of a new kind of anomalous (or sub-ballistic) Lieb-Robinson bound for an isotropic XY chain in a quasi-periodic transversal magnetic field. Instead of the usual effective light-cone $|x|\leq v|t|$, we obtain $|x|\leq v|t|^\alpha$ for some $0<\alpha<1$. We can characterize the allowed values of $\al$ exactly as those exceeding the upper transport exponent $\al_u^+$ of a one-body Schr\"odinger operator. To our knowledge, this is the first rigorous derivation of anomalous quantum many-body transport. We also discuss anomalous LR bounds with power-law tails for a random dimer field.
\end{abstract}

\maketitle



\section{Introduction}
Relativistic systems are local in the sense that information propagates at most at the speed of light. In their seminal paper \cite{LiebRobinson72}, Lieb and Robinson found that non-relativistic quantum spin systems described by local Hamiltonians satisfy a similar ``quasi-locality'' under the Heisenberg dynamics. Their \emph{Lieb-Robinson bound} and its recent generalizations \cite{HastingsLSM, NachtergaeleSims06} implies the existence of a ``light cone'' $|x|\leq v|t|$ in space-time, outside of which quantum correlations (concretely: commutators of local observables) are exponentially small. In other words, the LR bound shows that, to a good approximation, quantum correlations \emph{propagate at most ballistically}, with a system-dependent ``Lieb-Robinson velocity'' $v$.

About ten years ago, the general interest in LR bounds re-surged when Hastings and co-workers realized that they are the key tool to derive exponential clustering, a higher-dimensional Lieb-Schultz-Mattis theorem and the celebrated \emph{area law for the entanglement entropy} in one-dimensional systems with a spectral gap \cite{HastingsKoma06, HastingsLSM, HastingsAL}. These results highlight the role of entanglement in constraining the structure of ground states in gapped systems and yield many applications to quantum information theory, e.g.\ in developing algorithms to simulate quantum systems on a classical computer \cite{BHV, BHVC}.

In this paper, we announce and sketch the rigorous proof of a new kind of \emph{anomalous (or sub-ballistic) Lieb-Robinson bound} for an isotropic XY chain in a quasi-periodic transversal magnetic field. The LR bound is anomalous in the sense that the forward half of the ordinary light cone is changed to the region $|x|\leq v|t|^{\al}$ for some $0<\alpha<1$.

Previous study has focused on the dependence of the Lieb-Robinson velocity $v$ on the system details \cite{NachtergaeleSims06}, with particular interest in the case $v=0$, since it may be interpreted as dynamical localization \cite{HSS11}. In a very recent paper \cite{Gongetal}, a logarithmic light-cone was obtained for long-range, i.e.\ power-law decaying, interactions. The anomalous LR bound we find yields a \emph{qualitatively completely different}, anomalously slow many-body transport.

We expect that if one has an anomalous LR bound for a system with a spectral gap, the arguments of \cite{HastingsKoma06, NachtergaeleSims06} will yield anomalously strong exponential clustering (see the discussion after Def.\ 1).

We actually have an exact characterization of the values of $\al$ for which the anomalous LR bound holds, namely whenever $\al$ exceeds $\al_u^+$, the upper transport exponent of the one-body discrete Schr\"odinger operator with potential given exactly by the quasi-periodic field. Thanks to extensive study, there exist both rigorous and numerical upper and lower bounds on $\al_u^+$ \cite{AH, D05, DEG, DGY, DKL, DT07, DT08}.

We mention that quasi-periodic sequences serve as models for one-dimensional quasi-crystals and their sometimes exotic transport properties. Especially the discrete one-body Schr\"odinger operator with Fibonacci potential, see \eqref{eq:Hdefn}, has been considered \cite{KKT, OPRSS, AH, C, S, GA, S2, DL, DKL, DT07, DT08, DG, DEG, DGY}. Quasi-periodic spin chains (in particular with Fibonacci disorder) have also been studied extensively, with a focus on spectral properties and critical phenomena \cite{Benza1990a, Benza1990b, Benza1989, Doria1988, Igloi2007, Ceccatto1989, Satija1990, Luck1993}.

While we give the full statements below, we only give a rough sketch of the proof; a detailed version will appear elsewhere \cite{DLLY}.

\section{Setup and Main Result}
For any integer $N$, we consider the isotropic XY chain defined by the Hamiltonian
\beq
\label{eq:HNdefn}
    H_N= -\sum_{x=1}^{N-1} \l(\sigma_x^{1}\sigma_{x+1}^{1}+\sigma_x^{2}\sigma_{x+1}^{2}\r)
    + \sum_{x=1}^N h_x \sigma_x^{3},
\eeq
where $\sigma^1,\sigma^2,\sigma^3$ are the usual Pauli matrices. We scaled out the usual $J$ factor in front of the first term and chose zero boundary conditions for convenience. For definiteness, we let $h_x$ be the \emph{Fibonacci magnetic field}
\beq
\label{eq:fibonaccidefn}
h_x = \lam\chi_{[1-\phi,1)}(x\phi+\om \!\!\! \mod 1)
\eeq
where $\lam>0$ is a coupling constant, $\om\in[0,1)$ is an arbitrary phase offset, and $\phi$ is the inverse of the golden mean, i.e.
\beqs
\phi=\frac{\sqrt{5}-1}{2}.
\eeqs
The Fibonacci field \eqref{eq:fibonaccidefn} is prototypical in the study of one-dimensional quasi-crystals, but in fact $\phi$ can be replaced by an arbitrary irrational number in $(0,1)$ here (``Sturmian class''); compare \cite{KKT, OPRSS, SBGC, BIST, DEG}. We let $\curly{O}_x$ denote the set of observables at site $x$, which is of course just the set of Hermitian $2\times 2$ matrices, and for an observable $A$, we let
\beq
\label{eq:Heisenbergdefn}
A(t)\equiv e^{it H_N} A e^{-it H_N}
\eeq
be its image under the Heisenberg evolution after time $t$. Note that $A(t)$ implicitly depends on $N$ as well.

\begin{defn}[anomalous LR bound]
We say that $\mathbf{LR(\al)}$ holds if there exist positive constants $C,\xi,v$ such that for all integers $x,x',N$ with $1\leq x<x'\leq N$ and all times $t>0$, the bound
\beq
\label{eq:LRal}
    \|[A(t),B]\|\leq C \|A\| \|B\| e^{-\xi (|x-x'|-vt^{\alpha})}
\eeq
holds for all observables $A\in\curly{O}_x$ and $B\in \curly{O}_{x'}$.
\end{defn}
Let us make a few remarks about this: Firstly, the usual Lieb-Robinson bound corresponds to $\mathbf{LR(1)}$ and is known to hold by general considerations \cite{LiebRobinson72}. When comparing $\mathbf{LR(\al)}$ with $\mathbf{LR(1)}$ in the particularly relevant regime of small times, it is important to keep in mind that $|x-x'|\geq 1$ by definition and consequently $|x-x'|^{1/\al} > |x-x'|$ for $0<\al<1$. Hence, for fixed $t$, $\mathbf{LR(\al)}$ is effective at smaller distances than $\mathbf{LR(1)}$. Secondly, \eqref{eq:LRal} can be extended to a much wider class of observables, provided that their supports are a non-zero distance apart \cite{DLLY, DLY}. Thirdly, we emphasize that the constants above do not depend on the system size $N$, so that the estimate \eqref{eq:LRal} is stable in the thermodynamic limit $N\rightarrow \infty$. Finally, as mentioned in the introduction, if one can prove $\mathbf{LR(\al)}$ for a system with a spectral gap, we expect that ground-state correlations will decay anomalously fast, i.e.\ the usual exponential decay in $d(X,Y)$ is replaced by decay in $d(X,Y)^{1/\al}$ (see e.g.\ Theorem 2 in \cite{NachtergaeleSims06}). Essentially, this should follow from the proofs in \cite{HastingsKoma06, NachtergaeleSims06}, by using $\mathbf{LR(\al)}$ instead of $\mathbf{LR(1)}$, which only changes the optimization problem in the time cutoff parameter (called $s$ in \cite{NachtergaeleSims06}).

Our first main result is:

\be{thm}
Let $\lam\geq 8$. There exists $0<\al<1$ such that $\mathbf{LR(\al)}$ holds.
\e{thm}

As mentioned in the introduction, we actually have a characterization of the values of $\al$ for which $\mathbf{LR(\al)}$ holds for all $\lam>0$. This characterization is in terms of the upper transport exponent $\al_u^+$ of the one-body discrete Schr\"odinger operator $\mathfrak{h}$ with Fibonacci potential. It acts on a square-summable sequence $\{\psi_x\}_{x\geq 1}$ by
\beq
\label{eq:Hdefn}
 \l(\mathfrak{h}\psi\r)_x =\psi_{x+1}+\psi_{x-1}+ h_x \psi_x,
\eeq
with $\psi_0\equiv 0$ and $h_x$ given by \eqref{eq:fibonaccidefn}. $\al_u^+$ is then the propagation rate of the fastest part of an initially localized wave-packet. Since exponential tails cannot be evaded in quantum mechanics, $\al_u^+$ is, roughly, the largest exponent $\beta$ for which the probability of an initially localized wavepacket to travel a distance $t^{\beta}$ in time $t$ is \emph{not} exponentially small.

More formally: For any integer $x\geq 1$ and any positive real number $\beta$, let
\begin{align}
\label{eq:Pdefn}
    P(x,t)&=\sum_{x'>x} |\langle \de_{x'} \vert e^{-it\mathfrak{h}}\vert\de_1\rangle|^2,\\ R^+(\beta) & = - \limsup_{t\rightarrow \infty} \frac{\log P(t^\beta,t)}{\log t}. \label{eq:Rplusdefn}
\end{align}
Then, we define
\beq
\label{eq:aldefn}
\alpha^+_u = \sup_{\beta\geq 0}\l\{ R^+(\beta)<\infty\r\}.
\eeq
Note that $\al_u^+ = \al_u^+(\lam)$. We mention that $\al_u^+$ is just one of several transport exponents commonly associated to anomalous one-body dynamics \cite{DT07, DT08}, but as it turns out it is the only one relevant for LR bounds.

As anticipated before, we have the following characterization:

\begin{thm}
Let $\lam>0$. If $\al>\al_u^+$, then $\mathbf{LR(\al)}$ holds. Conversely, if $\al < \al_u^+$, then $\mathbf{LR(\al)}$ does not hold.
\end{thm}

In words, $\mathbf{LR(\al)}$ is a precise way to state that tails are exponentially decaying beyond a modified light-cone of the form $|x|\le v t^\alpha$, and our theorem states that this is true for $\alpha > \alpha_u^+$ and false for $\alpha < \alpha_u^+$. In fact, the second statement holds for completely general transversal magnetic fields (e.g.\ periodic ones, where $\al_u^+=1$). At first sight, it may be surprising that the quantity $\al_u^+$, which describes large-time asymptotics, characterizes the LR bound. Intuitively, this is due to the fact that the asymptotics capture precisely the fastest moving part of the one-body dynamics.

We also obtain an explicit expression for the LR velocity $v$, see (38) in \cite{DLLY}. Appropriately, $v$ is a decreasing function of $\al$.

 Let us discuss $\al_u^+$ from a quantitative viewpoint. Since Theorem~2 holds for arbitrary coupling constant $\lam>0$, we see that the restriction to $\lam\geq 8$ in Theorem~1 is due to the fact that we do not know rigorously that $\al_u^+<1$ for \emph{all} $\lam>0$ (we do know that $\al_u^+>0$ for all $\lam>0$ \cite{DKL}). We emphasize that estimating $\al_u^+$ is only a problem of \emph{one-body dynamics} however, which is simpler from both a theoretical and a numerical standpoint. A rough numerical study we conducted suggests that $\al_u^+<1$ also holds for $0\ll \lam<8$, and we think it would be interesting to pursue the numerical aspects further. Moreover, explicit rigorous upper and lower bounds for $\al_u^+$ exist \cite{DT07, DT08, DGY}. Asymptotically, they behave like $\frac{2 \log (1 + \phi)}{\log \lam}$ for large $\lam$ and they can be used to obtain quantitative estimates, such as
\beqs
	0.1<\al_u^+ < 0.5
\eeqs
for all $12 \leq \lam \le 7,000$. We stress the upper bound by $0.5$ because the particular case $\al_u^+ = 0.5$ is sometimes called diffusive transport and not assigned the ``anomalous'' label.

\section{Sketch of Proof}
Following \cite{LSM61}, we map the XY chain to free fermions via the Jordan-Wigner transformation. That is, we introduce the spin raising and lowering operators
\beqs
 S^{\pm} = \frac{1}{2}\l( \sigma^1\pm i \sigma^2  \r),
\eeqs
and define
\beq
\label{eq:JW}
    c_1= S^-_1,\qquad c_x = \sigma^3_1 \ldots \sigma^3_{x-1} S^-_x.
\eeq
These operators satisfy the CAR and allow us to rewrite the Hamiltonian as
\beqs
 H_N = \sum_{x=1}^{N} \sum_{y=1}^N c_x^\dag (\goth{h}_N)_{x,y} c_y.
\eeqs
Here, 
$\goth{h}_N$ is the operator $\goth{h}$ defined in \eqref{eq:Hdefn}, but with a zero boundary condition at site $N+1$. At this stage, $H_N$ can be diagonalized by a standard Bogoliubov transformation. One finds the following formula \cite{HSS11} for the Heisenberg dynamics \eqref{eq:Heisenbergdefn} of the fermion operators:
\beq
\label{eq:cjdynamics}
 c_x(t) =\sum_{y=1}^N \l(e^{-2i \goth{h}_N t}\r)_{x,y} c_y.
\eeq

\be{defn}
We say that $\mathbf{LR_{\mathrm{fermi}}(\al)}$ holds if there exist positive constants $C,\xi,v$ such that for all integers $x,x',N$ with $1\leq x < x'\leq N$ and all times $t>0$, the bound
\beq
\label{eq:LRalfermi}
    \|[c_x(t),B]\|+\|[c_x^\dag(t),B]\|\leq C\|B\| e^{-\xi (|x-x'|-vt^{\alpha})}
\eeq
holds for all observables $B\in \curly{O}_{x'}$.
\e{defn}

As we will see, \eqref{eq:cjdynamics} allows us to prove $\mathbf{LR_{\mathrm{fermi}}(\al)}$ by controlling the one-body transport created by $\goth{h}$. This is not surprising, because \eqref{eq:cjdynamics} is an expression of the fact that we are now describing free particles.

The problem that arises, though, is that the Jordan-Wigner transformation \eqref{eq:JW} is \emph{highly non-local}, while a Lieb-Robinson bound is of course an \emph{inherently local} statement. The key lemma, which is somewhat surprising at first sight, however says

\be{lm}
\label{lm:equiv}
$\mathbf{LR_{\mathrm{fermi}}(\al)}$ is equivalent to $\mathbf{LR(\al)}$.
\e{lm}

The point is that, as originally realized in \cite{HSS11} and adapted here to our purposes, inverting the non-local Jordan-Wigner transformation essentially just requires \emph{summing} up fermionic LR bounds: By an iteration argument, which is based only on $(AB)(t)=A(t)B(t)$ and the usual commutator rules, one can show
\beq
\label{eq:lmproof}
    \|[S_x^-(t),B]\| \leq 2\sum_{y=1}^x \l(\|[c_y(t),B]\|+\|[c_y^\dag(t),B]\|\r)
\eeq
for all $B\in \curly{O}_{x'}$. By taking adjoints and using commutator rules, similar bounds hold for $S_x^-,\, S_x^-S_x^+,\, S_x^+S_x^-$ and hence for all elements of the four-dimensional algebra of observables $\curly{O}_x$. Assuming that $\mathbf{LR_{\mathrm{fermi}}(\al)}$ holds,
we now see that $\mathbf{LR(\al)}$ follows from \eqref{eq:lmproof} and the trivial, but important, fact that
\beqs
\sum_{y=1}^{x} e^{-\xi(|y-x'|-vt^{\al})} \propto e^{-\xi(|x-x'|-vt^{\al})}.
\eeqs
For more details and the argument for the converse statement, see \cite{DLLY}. In conclusion, we found that the price of non-locality was the additional sum over $y$ in \eqref{eq:lmproof}, but we can afford this because \emph{tails of exponentially decaying series still decay exponentially}.

To prove Theorem 2, thanks to Lemma 1, it remains to characterize the values of $\al$ for which $\mathbf{LR_{\mathrm{fermi}}(\al)}$ holds. We first show that $\al>\al_u^+$ implies $\mathbf{LR_{\mathrm{fermi}}(\al)}$.
By \eqref{eq:cjdynamics} and the fact that $c_y$ and $B$ commute for $y<x'$, we get
\beq\label{eq:cxtestimate}
\|[c_x(t),B]\| \leq \|B\| \sum_{y=x'}^N \l| \langle\de_x \vert e^{-2i\goth{h}_Nt}\vert \de_y \rangle \r|.
\eeq
Since spatial translation corresponds to a shift of the (anyway arbitrary) phase offset $\om$, modulo some technical difficulties, the right-hand side is equal to
\beq
\label{eq:sumabove}
    \sum_{y=x'-x-1}^{N-x-1} \l| \langle\de_1 \vert e^{-2i\goth{h}_Nt}\vert \de_y \rangle\r|
\eeq
and this expression is already quite similar to the definition of the ``outside probability'' in \eqref{eq:Pdefn}. This explains why we can apply techniques developed in \cite{D05, DT07, DT08, DGY} to study the transport exponent $\al_u^+$ to our situation. A rough outline of the by now standard approach reads:

\be{itemize}
\item[(a)] use Dunford's formula
\beqs
    \langle\de_1 \vert e^{-2i\goth{h}_Nt}\vert \de_y \rangle = -\frac{1}{2\pi i} \int_\Gam e^{-itz}  \langle\de_1 \vert \frac{1}{-2\goth{h}_N-z}\vert \de_y \rangle\d z
\eeqs
to express the time-evolution in terms of resolvents ($\Gamma$ is a simple positively oriented contour around the spectrum of $-2\goth{h}_N$),
\item[(b)] bound matrix elements of resolvents in terms of transfer matrix norms, by studying individual solutions,
\item[(c)] bound transfer matrix norm by the exponentially decaying right-hand side in $\mathbf{LR_{\mathrm{fermi}}(\al)}$, by studying the Fibonacci trace map.
\e{itemize}

However, the original results of \cite{D05, DT07, DT08, DGY} do not translate directly to our situation. Firstly, the operator $\goth{h}$ lives on the half-line, while $\goth{h}_N$ has a zero boundary condition at $N+1$. This is a minor obstruction and can be removed, for an upper bound, by one-rank perturbation theory on the level of resolvents.

 The bigger problem is that the summands in \eqref{eq:sumabove} are \emph{not squared}, as they are in \eqref{eq:Pdefn}, which may of course make for a much larger sum. The technical solution we have found to this will not be presented here for the sake of brevity and instead we refer the interested reader to \cite{DLLY}.

We now turn to the converse direction in Theorem 2.
We prove the logically equivalent statement that $\mathbf{LR_{\mathrm{fermi}}(\al)}$ implies $\al \ge \al_u^+$. Using \eqref{eq:cjdynamics} and an appropriate trial state to bound the operator norm (see \cite{DLLY} for details), we obtain the key estimate
\[
\|[c_x(t),S^+_{x'}]\| \geq \l| \langle\de_x \vert e^{-2i\goth{h}_Nt}\vert \de_{x'} \rangle \r|
\]
(compare with \eqref{eq:cxtestimate}). Thus, $\mathbf{LR_{\mathrm{fermi}}(\al)}$ implies
\[
\l| \langle\de_x \vert e^{-2i\goth{h}_Nt}\vert \de_{x'} \rangle \r| \leq C e^{-\xi (|x-x'|-vt^{\alpha})}
\]
for all $1 \le x \le x' \le N$ and all $t>0$. We take the limit $N\to \infty$ to pass to the half-line operator,
\beq\label{eq:upperestimateonebody}
\l| \langle\de_x \vert e^{-2i\goth{h} t}\vert \de_{x'} \rangle \r| \leq C e^{-\xi (|x-x'|-vt^{\alpha})}
\eeq
for all $x, x' \in \mathbb{N}$ and all $t>0$.
Using this on definition \eqref{eq:Pdefn} gives
\[
P(t^\beta,t) \le \frac {C^2}{1-e^{-2\xi}} e^{-2\xi (t^\beta - v (t/2)^\alpha)} \le \tilde C e^{-\xi t^\beta }
\]
whenever $\beta>\alpha$. By definitions \eqref{eq:Rplusdefn}, \eqref{eq:aldefn} we conclude that $\beta \ge \alpha_u^+$, so $\alpha \ge \alpha_u^+$.

\section{The Random Dimer Model}
We explain why our method does not extend to yield an anomalous LR bound with \emph{power-law tails} for the random dimer model \cite{DWP}. The focus is on ideas here, for a detailed discussion see \cite{DLLY}.

Recall the one-body discrete Schr\"odinger operator $\goth{h}$ from \eqref{eq:Hdefn}. In the random dimer model, the potential $h_n$ is a random variable taking either of the two values $\pm \lam$, each with probability $1/2$ say, but these values must always occur in pairs (or dimers). The intuition, due to Anderson's work, that a one-dimensional disordered quantum system should exhibit localization is only almost correct here: There exist critical energies $E_c=\pm \lam$ for which the transfer matrices across dimers commute and the system shows anomalous transport. As it turns out, the anomalous transport is so fast that $\al_u^+=1$ and so we cannot hope for an $\mathbf{LR(\al)}$ with $\al<1$.

Intuitively, this is because $\al_u^+=1$ means that the probablity to find the particle within a distance $t^\beta$ of its initial location after time $t$, is \emph{not} exponentially small for $\beta<1$. However, in the random dimer model, this probability is \emph{polynomially} small for some $\beta<1$. In fact, there are similar transport exponents $\tilde \beta^+(p)$, related to time-averaged $p$-th moments of the position operator, which characterize when this is the case and which were determined explicitly in \cite{JS, JSS}.

With this in mind, one may hope to use our method to find an \emph{anomalous LR bound with power-law tails}, which would be of the general form
\beq
\label{eq:LRPL}
    \|[A(t),B]\|\leq  C\|A\| \|B\| \l(\frac{|t|^{\gam(p)}}{|x-x'|}\r)^{p}
\eeq
for any $p\geq 0$ and some $0<\gam(p)<1$, that is related to $\tilde \beta^+(p)$. A problem arises, however, when we want to ``pull back'' the LR bound through the Jordan-Wigner transformation, as we did to prove Lemma 1. As we explained, the non-locality gives rise to the extra sum in \eqref{eq:sumabove}. While we stressed that the sum was irrelevant in the case of exponential decay, \emph{power-law decay decreases by one order under summation} and it turns out that this restricts $\gam(p)$ in \eqref{eq:LRPL} to $\gam(p)>1$. Of course, the ordinary LR bound is then again a better estimate and the argument is inconclusive.

\section{Conclusions}
We have sketched the rigorous proof of anomalous Lieb-Robinson bounds \eqref{eq:LRal}
for isotropic XY chains with a quasi-periodic transverse field, which can be viewed as models for quasi-crystals. To our knowledge, this is the first derivation of anomalous quantum many-body transport.

The characterization of the correct exponent $\al$ in the anomalous LR bound \eqref{eq:LRal} as the \emph{one-body} transport exponent $\al_u^+$ yields rigorous quantitative bounds on it and opens the anomalous LR bound up to numerical study.

We also present the concept of an anomalous LR bound with power-law tails \eqref{eq:LRPL}. While our argument is inconclusive for the random dimer model, we understand exactly why it fails. In particular, it would yield power-law LR bounds for models with somewhat smaller values of the transport exponent $\tilde \beta^+(p)$, if such models exist.

\section{Acknowledgments}

D. Damanik was supported in part by NSF grant DMS--1067988, M. Lukic was supported in part by NSF grant DMS--1301582, and W. Yessen was supported by NSF grant DMS--1304287. The authors wish to thank a referee for raising some interesting questions.

\end{document}